\documentclass[twocolumn]{revtex4}
\usepackage{amsmath}
\usepackage{amssymb}
\usepackage[dvips]{graphicx}
\usepackage{dcolumn}
\usepackage{color}
\begin{document}

\title{Polariton $\mathbb{Z}$ Topological Insulator}
\author{A. V. Nalitov}
\author{D. D. Solnyshkov}
\author{G. Malpuech}

\affiliation{Institut Pascal, PHOTON-N2, Clermont Universit\'{e}, Blaise Pascal
University, CNRS, 24 avenue des Landais, 63177 Aubi\`{e}re Cedex, France.}

\begin{abstract}

Recent search for optical analogues of topological phenomena mainly focuses on mimicking the key feature of quantum Hall \cite{Klitzing1980} and quantum spin Hall \cite{Kane2005} effects (QHE and QSHE): edge currents protected from disorder. QHE relies on time-reversal symmetry breaking, which can be realised in photonic gyromagnetic crystals \cite{Haldane2008}. In the optical range, the weak magneto-optical activity may be replaced with helical design of coupled waveguides, converting light propagation into a time-dependent perturbation \cite{Rechtsman2013}. Finally, optical QHE due to artificial gauge fields \cite{Ozawa2014} was predicted in microcavity lattices. Here, we consider honeycomb arrays of microcavity pillars \cite{Jacqmin2014} as an alternative optical-frequency 2D topological insulator. We show that the interplay between the photonic spin-orbit coupling natively present in this system \cite{Leyder2007, Nalitov2014} and the Zeeman splitting of exciton-polaritons in external magnetic fields \cite{Fischer2014} leads to the opening of a non-trivial gap characterised by $C=\pm 2$ set of band Chern numbers and to the formation of topologically protected one-way edge states.

\end{abstract}

%\pacs{}
\maketitle

\begin{figure}[t]\label{fig1}
\includegraphics[scale=1.6]{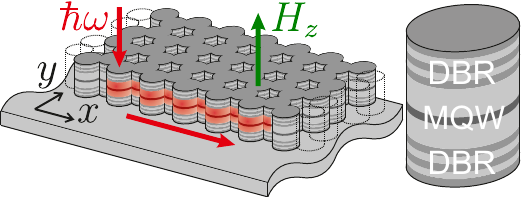} 
\caption{\textbf{Topologically protected light propagation through an edge of polariton topological insulator}.
The polariton graphene considered is based on an etched planar microcavity. The cavity is constituted by two Distributed Bragg Reflectors (DBRs) sandwiching a cavity with embedded Quantum Wells (QWs). The energy splitting existing between TE and TM polarised modes provides the photonic SOC. The application of a real magnetic field perpendicular to the x-y plane of the structures opens the non trivial gap. Edge modes are one way propagative modes which cannot elastically scatter to the bulk states. In a stripe geometry, normally incident light is guided either clockwise or anti-clockwise, depending on the external magnetic field sign. }
\end{figure} 

The history of topological insulators (TI) dates back to the discovery of QHE by Klaus von Klitzing in 1980 \cite{Klitzing1980}. In his experiment, a strong magnetic field pinned the conduction band electrons to the Landau levels, opening a band gap in the bulk and thus converting an electron conductor to an insulator. The edge electron states, on the contrary, carried one-way currents, protected from backscattering and responsible for the integer Hall conductance \cite{TKNN1982}. The classification of insulators, which allowed to distinguish this phase from a conventional band insulator, is based on the Chern topological invariant \cite{Simon1983} -- an integer number characterizing the band structure in terms of Berry phase.

A whole new family of TI materials with different sets of topological invariants and symmetries were later proposed and discovered \cite{Kane2010}. Graphene has a special place in this family: it allowed the observation of QHE at room temperature \cite{Novoselov2007}, played a role of a model system for QHE without net magnetic flux \cite{Haldane1988} and the QSHE \cite{Kane2005}. The latter is associated with topologically-protected boundary spin currents and is characterised by a non-zero $\mathbb{Z}_2$ invariant stemming from the spin-orbit coupling (SOC) for electrons \cite{Kane2005b}. Although the extremely small SOC has not allowed to observe QSHE in graphene, it was later demonstrated in various 2D and 3D structures \cite{Konig2007,Hsieh2008}. A Floquet TI having a topologically nontrivial gap was realised in 2D heterostructure under a microwave-range electromagnetic irradiation \cite{Lindner2011}.

Many promising implementations of topological phases in bosonic systems were recently proposed in honeycomb photonic gyromagnetic waveguides \cite{Haldane2008,Wang2008}, coupled microcavities \cite{Carusotto2012}, coupled cavity rings \cite{Hafezi2011}, coupled waveguides with a spatial modulation \cite{Rechtsman2013}, and photonic waveguides (out of the optical range) based on metamaterials with bi-anisotropic behaviour \cite{Khanikaev2013,Chen2014}.

Cavity polaritons result from the strong coupling between confined cavity photons and QW excitons. They are photonic states, but with an exciton fraction, making them strongly interacting  -- a feature at the heart of polariton Bose Einstein Condensation and of the quantum fluid behavior of a polariton gas \cite{Carusotto2013b}. In this manuscript, we exploit other polariton features: the exciton (and thus polariton) Zeeman splitting \cite{Fischer2014}, and the polarisation splitting of photonic modes, interpreted as an effective SOC for polaritons \cite{Leyder2007, Nalitov2014, Sala2014}. The possibility of creating in-plane potentials for photons \cite{Kim2011,CerdaMendez2013,Jacqmin2014} is also vital. All these ingredients give the polariton platform a unique flexibility to engineer the photonic properties of structures at optical frequencies.

In this letter, rather than creating artificial gauge fields or using weak gyromagnetic optical activity to break the time-reversal symmetry, we propose to exploit the natural susceptibility of microcavity polaritons to the magnetic field and the effective SOC acting on polaritons in photonic nanostructures. We consider polaritons in a honeycomb potential, called polariton graphene \cite{Jacqmin2014}. We demonstrate that a real magnetic field applied along the growth axis allows the formation of an original $\mathbb{Z}$ topological insulator with Chern numbers $C=\pm 2$.  We find the protected edge states by tight-binding calculation of the honeycomb microcavity stripe eigenstates. This result is confirmed by direct numerical simulations.

\emph{Tight-binding model} 
A state of the polariton graphene can be described by a bispinor $\Phi = \left( \Psi_A^+, \Psi_A^-, \Psi_B^+, \Psi_B^- \right)^{\mathrm{T}}$, with $\Psi_{A(B)}^\pm$ -- the wave functions of the two sublattices and two spin components.
On such a basis, the effective Hamiltonian in the presence of a real magnetic field applied along the z-direction reads: 

\ref{Hamiltonian}):
\begin{equation}\label{Ham_H}
\mathrm{H}_\mathbf{k} = \left( \begin{matrix}
\Delta \sigma_z & \mathrm{F}_{\mathbf{k}} \\
\mathrm{F}_{\mathbf{k}}^\dagger & \Delta \sigma_z
\end{matrix} \right), \quad
\Delta = \vert x \vert ^2 g_X \mu_B H_z/2,
\end{equation}
where $x$ is the excitonic Hopfield coefficient, $g_X$ is the effective g-factor for the 2D exciton, $\mu_B$ is the Bohr magneton, and $H_z$ is the applied magnetic field, giving rise to polariton Zeeman splitting $\Delta$.

 \begin{equation} \label{Hamiltonian}
\mathrm{F}_{\mathbf{k}} = - \left( \begin{matrix}
f_{\mathbf{k}} J & f_{\mathbf{k}}^+ \delta J \\
f_{\mathbf{k}}^- \delta J & f_{\mathbf{k}} J
\end{matrix} \right),
\end{equation}
where complex coefficients $f_{\mathbf{k}}$,$f_{\mathbf{k}}^\pm$ are defined by:
\begin{equation}
f_{\mathbf{k}}=\sum_{j=1}^3 \exp(-\mathrm{i}\mathbf{k d}_{\varphi_j}),\quad
f_{\mathbf{k}}^\pm = \sum_{j=1}^3 \exp(-\mathrm{i}\left[\mathbf{k d}_{\varphi_j} \mp 2 \varphi_j \right]), \notag
\end{equation}
and $\varphi_j = 2 \pi (j-1) / 3$ is the angle between horizontal axis and the direction to the $j$th nearest neighbor of a type-A pillar.
$J$ is the polarisation independent tunneling coefficient, whereas $\delta J$ is the SOC-induced polarisation dependent term. Without the magnetic field, the Hamitonian (\ref{Ham_H}) can be exactly diagonalised \cite{Nalitov2014}. The energy dispersions obtained are relatively close to those of bilayer graphene %\cite{McCann2006},
and of a monolayer graphene in presence of Rashba SOC. 
%\cite{Rakyta2010}.
The polarisation texture of the eigenstates is, however, different. 

\begin{figure} \label{fig2}
\includegraphics[scale=0.31]{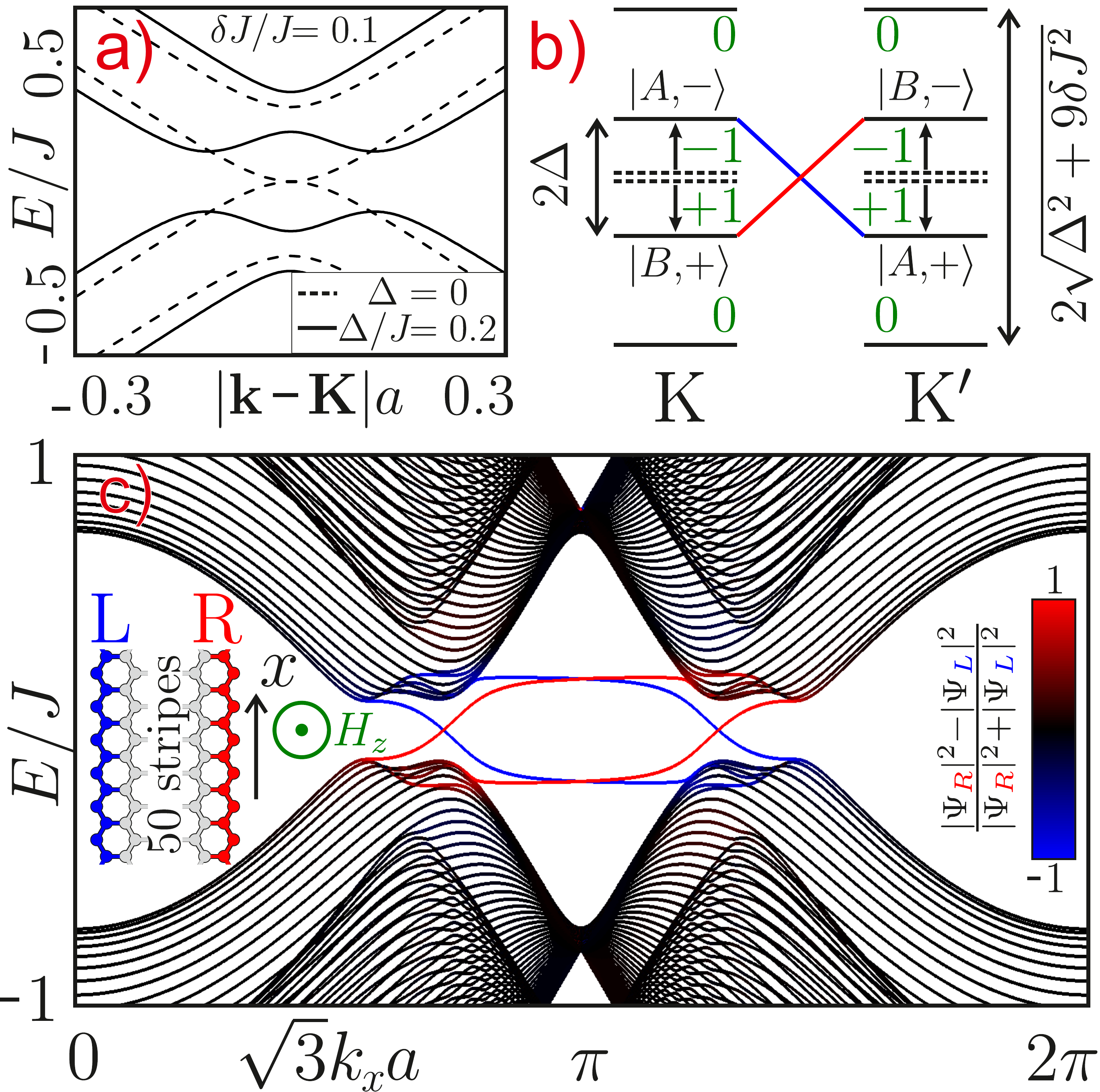} 
\caption{\textbf{Non-trivial band structure of the polariton graphene stripe in an external magnetic field.}
 a) bulk energy dispersion without (dashed) and with (solid) magnetic field. b) illustration of degeneracy lifting in K and K$^\prime$ points. Due to coupling between sublattice and polarisation, states, localised on one sublattice go up in energy at one point and down in the other. Chern numbers at each point are shown in green.  Bottom: numerical calculation of eighenstates: edge states are marked with colour. Direction of their propagation is set by the sign of the product $H_z g_X$ and is protected from both backscattering and scattering into the bulk.}
\end{figure}

The dispersion close to the $\mathbf{K}$ point is shown in dashed line on the figure (2-a). Under the effect of SOC, the Dirac point transforms into four inverted parabolas. Two parabolas are split off, while the two central ones cross each other. It is instructive to consider the eigenstates exactly at the Dirac points. At the $K$ point, the eigenstates of the two central parabolas are fully projected on $\Psi_A^-$ and $\Psi_B^+$ respectively, whereas at the $K'$ point they project on  $\Psi_A^+$ and $\Psi_B^-$ respectively. Let us now qualitatively consider the consequence of a finite Zeeman splitting. As sketched on the figure (2-b), a gap opens. At the $K$ point, the "valence" band is formed from the B-atoms and the "conduction" band from the A-atoms. On the contrary, at the $K'$ point, the valence band is formed from the A-atoms and the conduction band from the B-atoms. Therefore, the order of the bands is reversed (in the basis of the sublattices), which signifies a topological non-triviality of the gap\cite{Kane2010}. In the spin basis, however, the valence and conduction bands are equivalent at $K$ and $K'$, unlike the $\mathbb{Z}_{2}$-topological insulator \cite{Kane2010}. 

The result of the complete diagonalisation of the Hamiltonian (\ref{Ham_H}) is plotted with solid lines in the Figure 2-a. As expected, it shows an energy gap, saturating to $E_g \sim 3 \delta J$ at $\vert \Delta \vert \sim \delta J$. Without SOC ($\delta J=0$), the application of a magnetic field does not not open any gap. The band structure in this latter case is constituted by two graphene dispersions shifted in energy by the polariton Zeeman splitting.
The Chern numbers are numerically calculated from the Berry connection over the Brillouin zone \cite{Kane2010}:
\begin{equation}
n_m = {1 \over 2 \pi} \iint \limits_{\mathrm{BZ}} \mathbf{B}_{\mathbf{k},m} \mathrm{d}^2 \mathbf{k},
\end{equation}
where the Berry curvature $\mathbf{B}_{\mathbf{k},m}$ is expressed in the effective Hamiltonian (\ref{Ham_H}) and its eigenstates $\vert \Phi_{\mathbf{k},m} \rangle$ with corresponding energies $E_{\mathbf{k},m}$:
\begin{equation}
\mathbf{B}_{\mathbf{k},m} = i \sum_{l \neq m} 
\frac{\langle \Phi_{\mathbf{k},m} \vert \boldsymbol{\nabla}_\mathbf{k} \mathrm{H}_{\mathbf{k}} \vert \Phi_{\mathbf{k},l} \rangle \times
\langle \Phi_{\mathbf{k},l} \vert \boldsymbol{\nabla}_\mathbf{k} \mathrm{H}_{\mathbf{k}} \vert \Phi_{\mathbf{k},m} \rangle }{(E_{\mathbf{k},m}-E_{\mathbf{k},l})^2}.
\end{equation}
Two inner branches, split by the interplay of external magnetic field and effective SOC, have non-zero Berry connections around K and K$^\prime$ point, each giving $\pm 1$ contribution to the total band Chern number $\pm 2$ (marked in green in Fig. 2(b)).
Outer branches, on the contrary, have zero Berry curvature over all reciprocal space.

As a consequence of bulk-boundary correspondence, a finite micropillar honeycomb structure has one-way propagating edge states. 
To demonstrate this, we use the same tight-binding approach to model a quasi-1D stripe of microcavity pillars (see Methods), consisting of 50 zig-zag chains.
Figure 2c shows the result of the band structure calculation, where two pairs of edge states crossing the gap are marked with red and blue colours, corresponding to right and left edges (see inset).
The propagation direction of these edge states is related to the direction of the external magnetic field: the photon edge current is either clockwise or anti-clockwise depending on the signs of $H_z$ and $g_X$. One should insist on the fact that we deal with a real polariton current and not with a spin current.

In order to demonstrate the feasibility of experimental observations and to confirm our predictions, we carry out a full numerical simulation, describing the time evolution of the polariton wavefunction by solving the spinor Schrodinger equation:

\begin{eqnarray}
& i\hbar \frac{{\partial \psi _ \pm  }}
{{\partial t}}  =  - \frac{{\hbar ^2 }}
{{2m}}\Delta \psi _ \pm   + U\psi _ \pm   - \frac{{i\hbar }}
{{2\tau }}\psi _ \pm   \pm\Delta\psi _ \pm \\
& + \beta {\left( {\frac{\partial }{{\partial x}} \mp i\frac{\partial }{{\partial y}}} \right)^2}{\psi _ \mp } 
+\sum_{i}P_{i\pm} e^{ { - \frac{{\left( {t - t_0 } \right)^2 }}
{{\tau _0^2 }}}}e^{ { - \frac{{\left( {{\mathbf{r}} - {\mathbf{r}}_i } \right)^2 }}
{{\sigma ^2 }}}}e^{ {i\left( {{\mathbf{kr}} - \omega t} \right)} } \notag
\end{eqnarray}

where $\psi(\mathbf{r},t)={\psi_+(\mathbf{r},t), \psi_-(\mathbf{r},t)}$ are the two circular components of the wave function, $m$ is the polariton mass, $\tau$ the lifetime.

 \begin{figure} \label{fig3}
\includegraphics[scale=0.75]{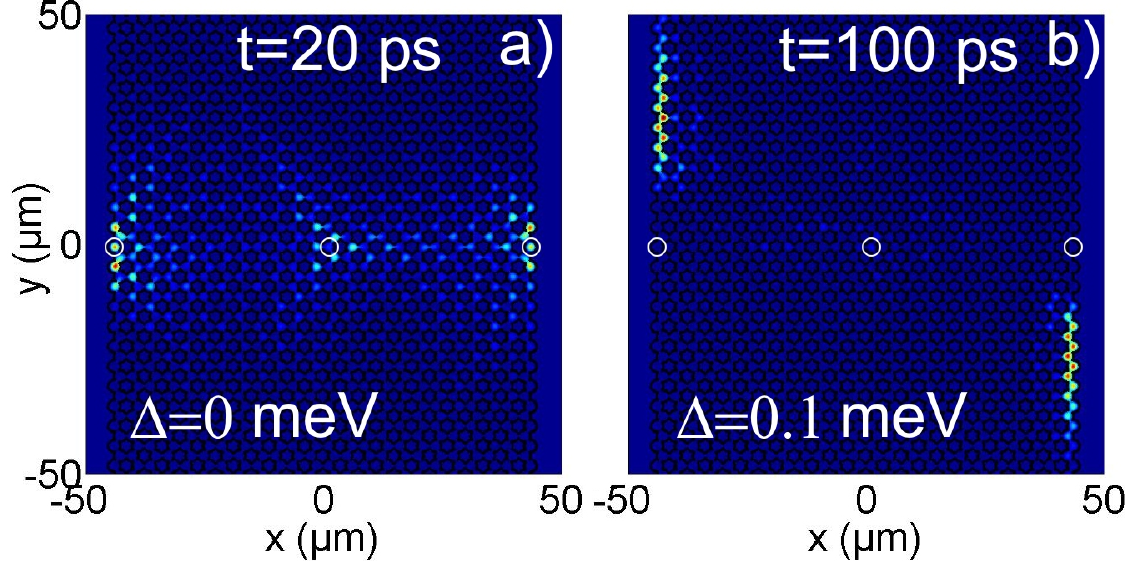} 
\caption{\textbf{Propagation of light in the conducting $(\Delta=0)$ and topological insulator phase $(\Delta\neq 0)$
Calculated spatial distribution of emission intensity.}
 a) rapid expansion of the bulk propagative states after 20 ps at $\Delta$=0; b) surface states after 100 ps at $\Delta$=0.1 meV. White circles show the pumping spots, and black line traces the contours of the potential. The parameters are: $\beta ={\hbar ^{2}}\left( {m_{l}^{-1}-m_{t}^{-1}}\right) /4m$ where $
m_{l,t}$ are the effective masses of TM and TE polarised particles
respectively and $m=2\left( {{m_{t}}-{m_{l}}}\right) /{m_{t}}{%
m_{l}}$; $m_t=5\times10^{-5}m_0$, $m_l=0.95m_t$, where $m_0$ is the free electron mass.$\tau_0=35$ ps, $\sigma=1$ $\mu$m $\omega=1.6$ meV, $\tau=25$ ps. Pumping $P$ was circular polarized.}
\end{figure}

Figure 3 demonstrates qualitatively different behaviour with and without the magnetic field. Without the field ($\Delta=0$), the excitation energy corresponds to propagating states, and the resulting expansion of polaritons is visible on panel b). However, under an applied field giving $\Delta=0.1$ meV, the excitation energy lies within a gap, which makes the injection in the center ineffective. But the spots on the edges become resonant with the surface states, and their one-way propagation is visible in panel c). The transverse profile of these states shows an exponential decay \cite{suppl} with a characteristic length of $\kappa=3.1\pm0.1~\mu$m, corresponding to the estimate $2\kappa\approx\sqrt{2mE_g/\hbar^2}$.
Thus, the full numerical simulation confirms the predictions of the tight-binding model on the appearance of the gap in the presence of magnetic field (see \cite{suppl} for dispersions) and the formation of one-way surface states.

All the parameters used in numerical simulations are entirely realistic. Indeed, the
experimental realisation of the effect has one important requirement: both $\delta J$ and $\Delta$ must exceed the broadening, which is of the order of $1/(25 \mathrm{ps})\approx 30 \mu$eV \cite{Jacqmin2014}.
The photonic SOC finds its origin in the polarisation splittings of photonic nanostructures. In etched planar cavities, it is induced by the TE-TM splitting \cite{Shelykh2010}, but also by strain and other structural effects which enhance s splittings up to 50-200 $\mu$eV in 1D ridges \cite{Dasbach2005}, or coupled pillar structures \cite{Galbiati2012}.  The Zeeman splitting between the spin components of polaritons can be of the order of 100-200 $\mu$eV at moderate magnetic fields (about 10 T) \cite{Fischer2014}.

In summary, we have demonstrated that a polariton graphene under magnetic field becomes a $\mathbb{Z}$ topological insulator with protected edge states at optical frequencies. It resembles the QHE behaviour of electrons and photons \cite{Carusotto2012}, both based on the appearance of Landau levels. However, its origin is different, being closer to the Haldane proposals \cite{Haldane1988,Haldane2008}. The gap appears in our case due to the Zeeman splitting of electrically uncharged particles and specific spin orbit coupling  acting together similarly to spatially alternating magnetic fields \citep{Haldane1988,Haldane2008}. The difference is underlined by the unique set of band Chern invariants $C_m=\pm2$ (transforming into $C_m=\pm1$ above critical SOC \cite{suppl}). From the point of view of symmetry, the topologically nontrivial gap is a manifestation of broken time-reversal and in-plane rotational symmetry by external magnetic field and effective SOC respectively. The interacting nature of polaritons opens interesting 
possibilities of studying collective bosonic effects \cite{Carusotto2013b} in  TIs. The spin-anisotropy of these interactions \cite{Shelykh2010} leads to self-induced Zeeman splitting, allowing a self-induced TI for a polarised polariton Bose-Einstein condensate.

We acknowledge discussions with M. Glazov, A. Amo, and J. Bloch. This work has been supported by the ITN INDEX (289968), ANR Labex GANEX (207681) and IRSES POLAPHEN (246912).

\section{Methods}

To construct the tight-binding Hamiltonian we first consider a
photonic molecule consisting of two coupled pillars and find its 4x4
Hamiltonian in the following basis of states $\vert A,L \rangle$,
$\vert A,T \rangle$, $\vert B,L \rangle$, $\vert B,T \rangle$.
Here $A/B$ define the pillar at which a state is localised, while
$L/T$ name polarisation of a state - either longitudinal or transverse
to $AB$ axis connecting the pillars.
Due to the circular symmetry of the pillars and their identity, all
basis states are degenerate, we set their energy to 0.
In absence of external magnetic field, within each pillar linearly
polarised states are uncoupled, therefore we find that 2x2 diagonal
blocks of the Hamiltonian consist of zeros. LT splitting acts on a propagating polariton as an effective field
with absolute value proportional to squared momentum, thus producing
two effective masses $m_L < m_T$ for states, polarised longitudinally
and transversely to propagation direction. We estimate potential barrier due to size quantisation at narrow
junction between the pillars as $V_{L(T)}=\pi^2 \hbar^2 / 2 m_{L(T)}
w^2$, where $w$ is the junction width. Note that the barrier value depends on the mass in direction,
orthogonal to $AB$ axis. For the state $\vert A,L(T) \rangle$ wavefunction tail at pillar $B$
is determined by tunnel extinction coefficient $\kappa_{L(T)}=\sqrt{2
m_{L(T)}(V_{T(L)}-E)}$.Therefore we find that the nondiagonal blocks have different matrix
elements $-J_{L,T}$, $J_L>J_T$ at their diagonals, whose values depend
on overlaps of corresponding wavefunctions.
Their nondiagonal terms are zero due to the fact that linearly
polarised eigenstates of LT effective field are not rotated during
propagation.
A transfer to the basis of circularly polarised states gives
$-J=-(J_L+J_T)/2$ matrix elements at diagonals and introduces
off-diagonal terms $-\delta J = -(J_L-J_T)/2$ .

Starting from this Hamiltonian we construct the polaritonic graphene
effective Hamiltonian on the basis of Bloch wavefunctions
$\Psi_{A/B}^{\pm}(\mathbf{k})$ in closest neighbour hopping
approximation. We account for magnetic field via Zeeman splitting $\Delta$ of states,
localised at a pillar, and finally arrive to Hamiltonian
(\ref{Ham_H}).
To demonstrate one-way edge states in tight-binding approach, we
derive a $4N$x$4N$ Hamiltonian for a polariton graphene tape,
consisting of $N$ infinite zig-zag stripes.
For this we set a basis of Bloch waves $\Psi_{A/B,n}^\pm(k_x)$, where
$n$ index numerates stripes, and $k_x$ is the quasi-wavevector in the
infinite zig-zag direction.
4x4 diagonal blocks describe coupling withing one stripe and are
derived in the same fashion as Hamiltonian (\ref{Ham_H}), coupling
between stripes is accounted for in subdiagonal 4x4 blocks.

The nVidia CUDA graphical processor was used to carry out the numerical integration of the 2D spinor Schroedinger equation. The high-resolution (1024x1024) honeycomb lattice potential $U(r)$ contains 23x30 elementary cells.

\section{Supplemental Material}

\subsection{Dispersion}
Similar to the tight binding model, the numerical analysis starts from the calculation of the dispersion. For this, a single excitation spot of a small size $\sigma=0.7~\mu$m is used, combined with a short pulse duration of $\tau_{0}=0.7$ ps, in order to excite a large part of the dispersion. Then, the Schrodinger equation (5) is integrated for 200 ps, and the resulting wavefunction $\psi(\mathbf{r},t)$ is Fourier-transformed over time and spatial coordinates to give $\psi(\mathbf{k},\omega)$, which is the dispersion. Such calculation was carried out with the excitation taking place in the "volume" of the polariton graphene structure and on its surface. The surface states were excited in the latter case.

The results of the calculations are shown in figure S1. Panel a) demonstrates the dispersion calculated when the pump spot is positioned on a pillar located in the bulk. In this case the surface states are not excited, and the gap opening due to the applied magnetic field is clearly visible aroung 1.6 meV. The image represents a cut of the 2D dispersion along the K'MK line, which allows to see two Dirac cones at the same time. Panel b) shows the results of the calculation when the pump spot is located on a pillar on the surface. In this case, both surface and bulk states are excited, and both are visible in the resulting dispersion. However, it is possible to distinguish the surface states since only these states can appear in the gap (marked by the dashed lines).

\begin{figure} \label{figS1}
\includegraphics[scale=0.6]{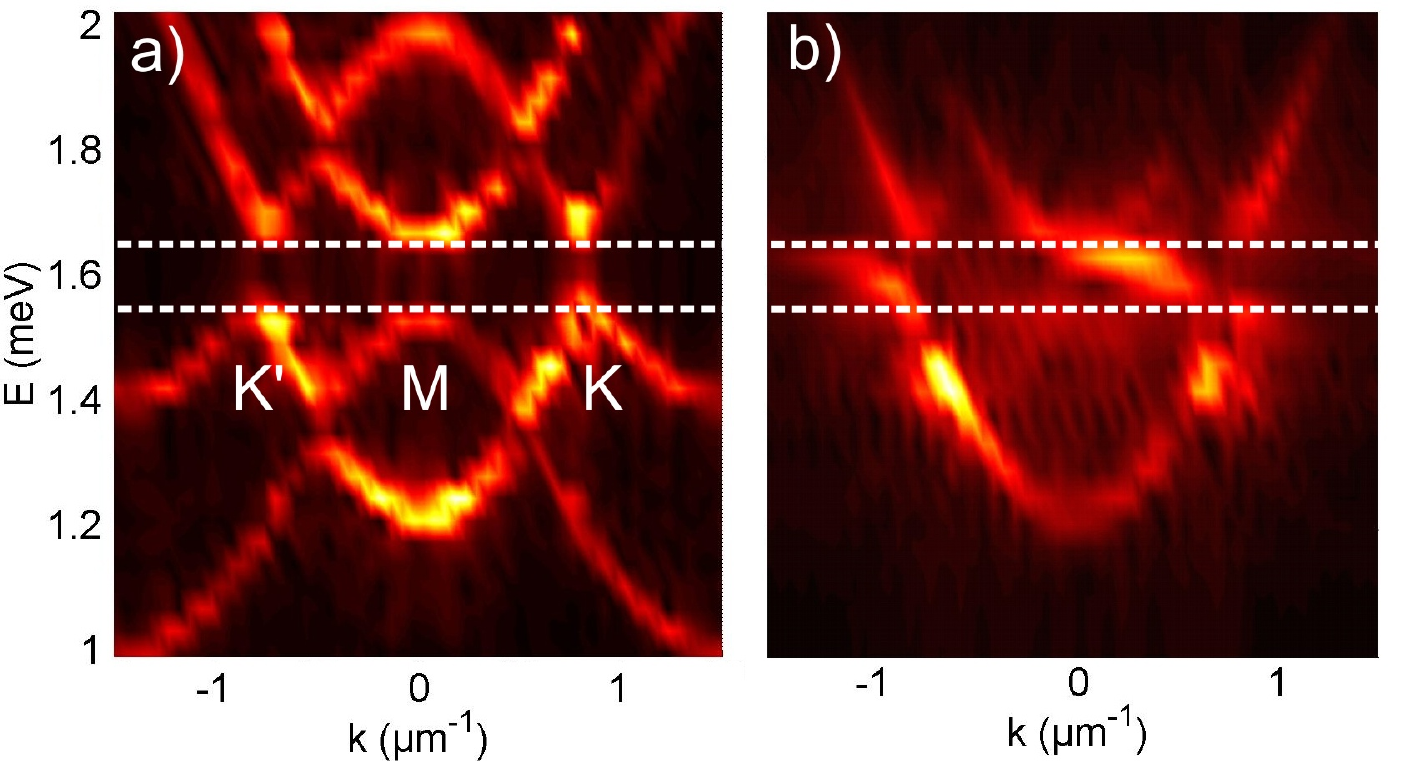} 
\caption{Numerically calculated dispersion of the polariton graphene. a) Bulk dispersion (cut across the K'MK edge of the 1st Brillouin zone). The gap opening due to the magnetic field is shown with the dashed white lines. b) Surface states dispersion. Two uni-directional states are visible in the gap of the bulk.}
\end{figure} 

\subsection{Profile of the surface states}

We have also analyzed the transverse profile of the surface states appearing under the effect of the magnetic field in the numerical simulations, as shown in figure 3b of the main text. This profile is plotted in figure S2 with a solid line, while the dashed line is the exponential decay fit with a characteristic density decay length of 3.1 $\mu$m. This value corresponds very well to the one expected from the analytical estimate based on the particle effective mass and the width of the gap.

\begin{figure} \label{figS2}
\includegraphics[scale=0.2]{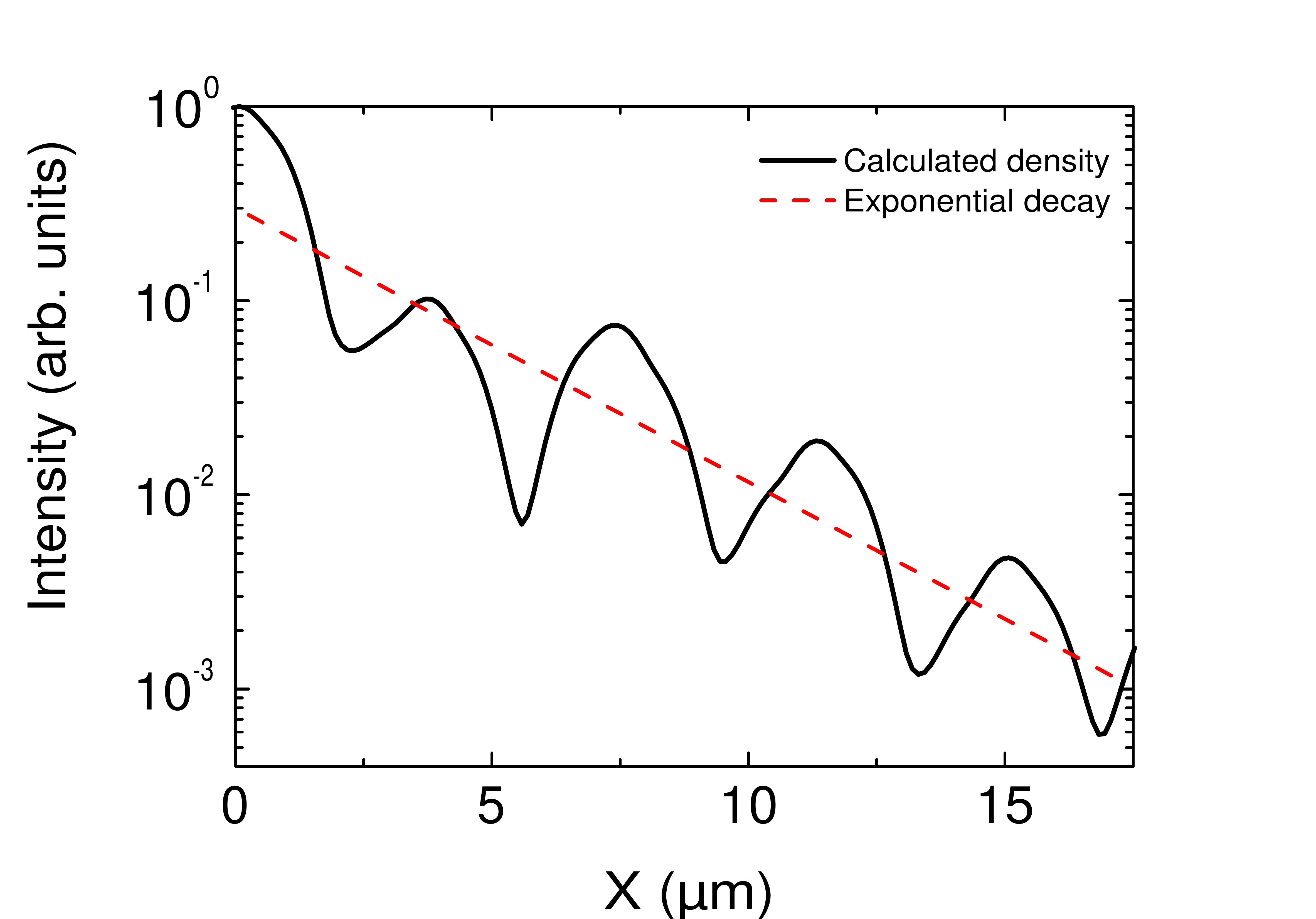} 
\caption{Transverse profile of the surface state. Black solid line - density profile obtained from numerical simulations; dashed red line - exponential fit.}
\end{figure}

\subsection{Analysis of the gap topology}

In this section, we propose and carry out numerically the experiments demonstrating the topological nature of the gap. Indeed, the tight-binding model predicts, that the nature of the "valence" and "conduction" bands is different in the K and K' points of the reciprocal space. In K point under magnetic field, the lower state is B+ and the upper state is A-, while in the K' point the lower state is A+ and the upper is B-.

This difference can be evidenced experimentally thanks to the possibility of direct optical measurements of the quantum states and also the controllable direct quasi-resonant excitation of the states of interest. For this, we have excited the system with a large pumping spot $\sigma=15~\mu$m, centering the pump first at the K point, and then at the K' point ($k=\pm1.1~\mu$m), keeping its energy $\hbar\omega=1.6$ meV within the gap. The pulse duration of $\tau_0=1.5$ ps is longer than for the dispersion calculation, but shorter than in the main text, in order to excite the states close to the gap. After the integration of the spinor Schrodinger equation, we make a Fourier transform of the wavefunction over time, and show the images corresponding to two given energies: those of the lower and upper states at the K or K' points.

The results are shown in figure S3. Panels a),c) were obtained exciting at the K point. They show the circular polarized states. The spin-down state ($\sigma^-$) has higher energy ("conduction" band) and is localized on the A-atoms (the atoms on the left of a horizontal pair A-B), as seen on panel a). The spin-up state ($\sigma^+$) has lower energy ("valence" band) and is localized on the B-atoms (the atom on the right of a horizontal pair A-B), as seen on panel c). Panels b),d) were plotted for the excitation at the K' point. The configuration here is opposite to that of K point, which demonstrates the topologically non-trivial nature of the gap. Here, the spin-down state ($\sigma^-$) ("conduction" band) is localized on the B-atoms in panel b) (it was A for panel a). At the same time, the spin-up state ($\sigma^+$, "valence" band, panel d) is localized on the A-atoms (while it was B for panel c). The arrows show how the surface states form from the bulk, as in fig. 2 of the main text.

\begin{figure}[t] \label{figS3}
\includegraphics[scale=0.75]{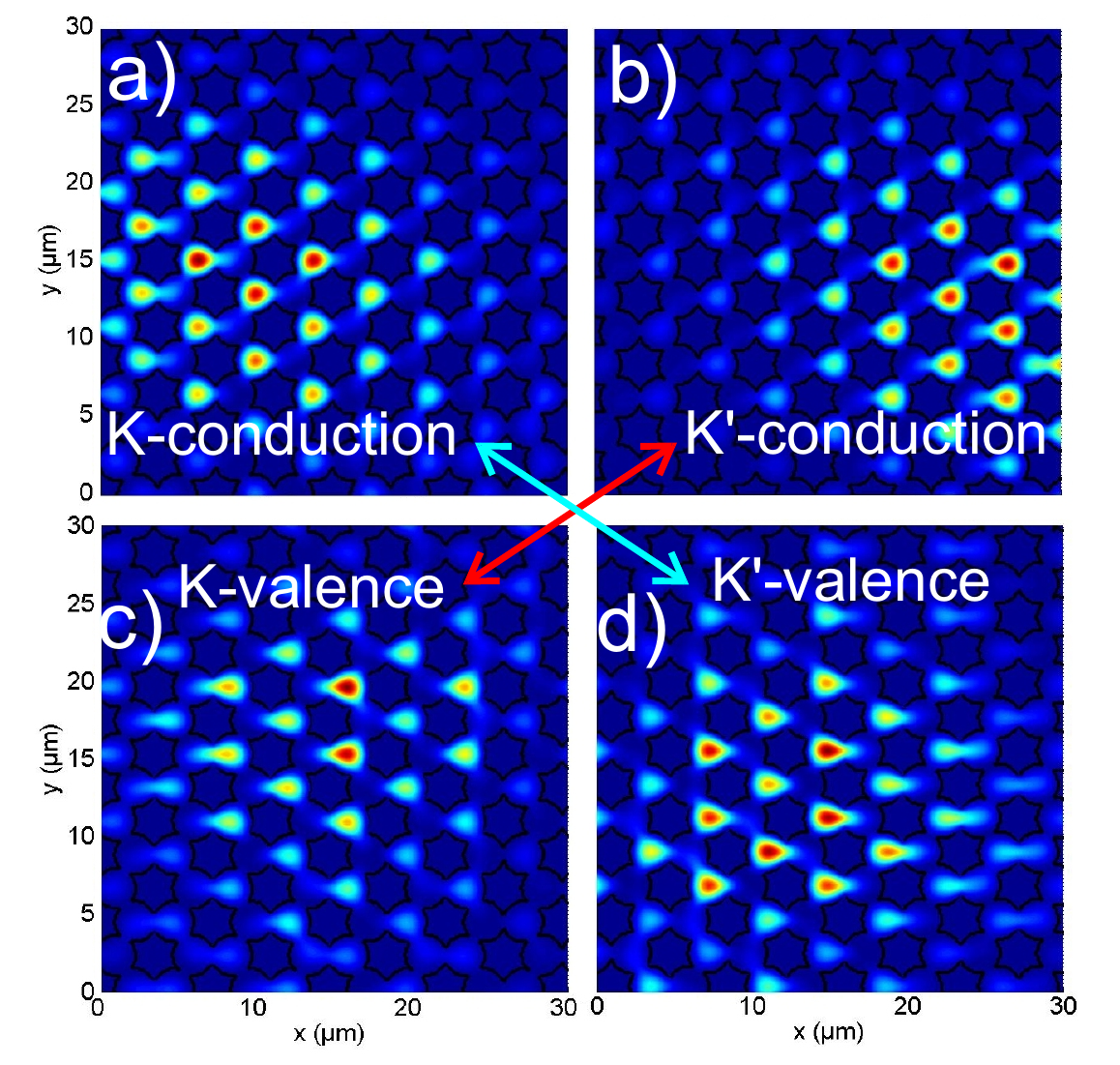} 
\caption{Real-space images corresponding to excitation in points K and K'. a) "conduction" band, K point; b) "conduction" band, K' point; c) "valence" band, K point; d) "valence" band, K' point. The arrows indicate the formation of the surface states. }
\end{figure} 

\subsection{Video}

The Supplemental videos 1\footnote{https://www.youtube.com/watch?v=6Xdzgke4R4I} and 2\footnote{https://www.youtube.com/watch?v=v8gF3of6BrA} correspond to the figure 3 of the main text. They demonstrate the time evolution of the polariton density distribution in real space: the propagation of the injected particles in the polariton graphene potential. Video 1, calculated without the magnetic field, corresponds to Fig3a). No gap is opening at the Dirac points, and therefore the pump is exciting the propagative states, which spread very rapidly. Video 2 corresponds to Fig3b, obtained under an applied magnetic field, which opens a gap in the bulk. The pumping spot in the center is not resonant with any states, and the injection becomes ineffective. At the same time, the pump spots on the surface inject particles into the one-way propagative surface states, which become clearly visible at later moments of time. The dispersion of surface states is much less steep than that of the bulk, and their group velocity is much smaller, which is why they take much longer time to propagate away from the pumping spots. If the sign of the magnetic field is inverted, the same configuration leads to the propagation in the opposite direction.

\subsection{Topological transition.}

\begin{figure} \label{figS4}
\includegraphics[scale=0.4]{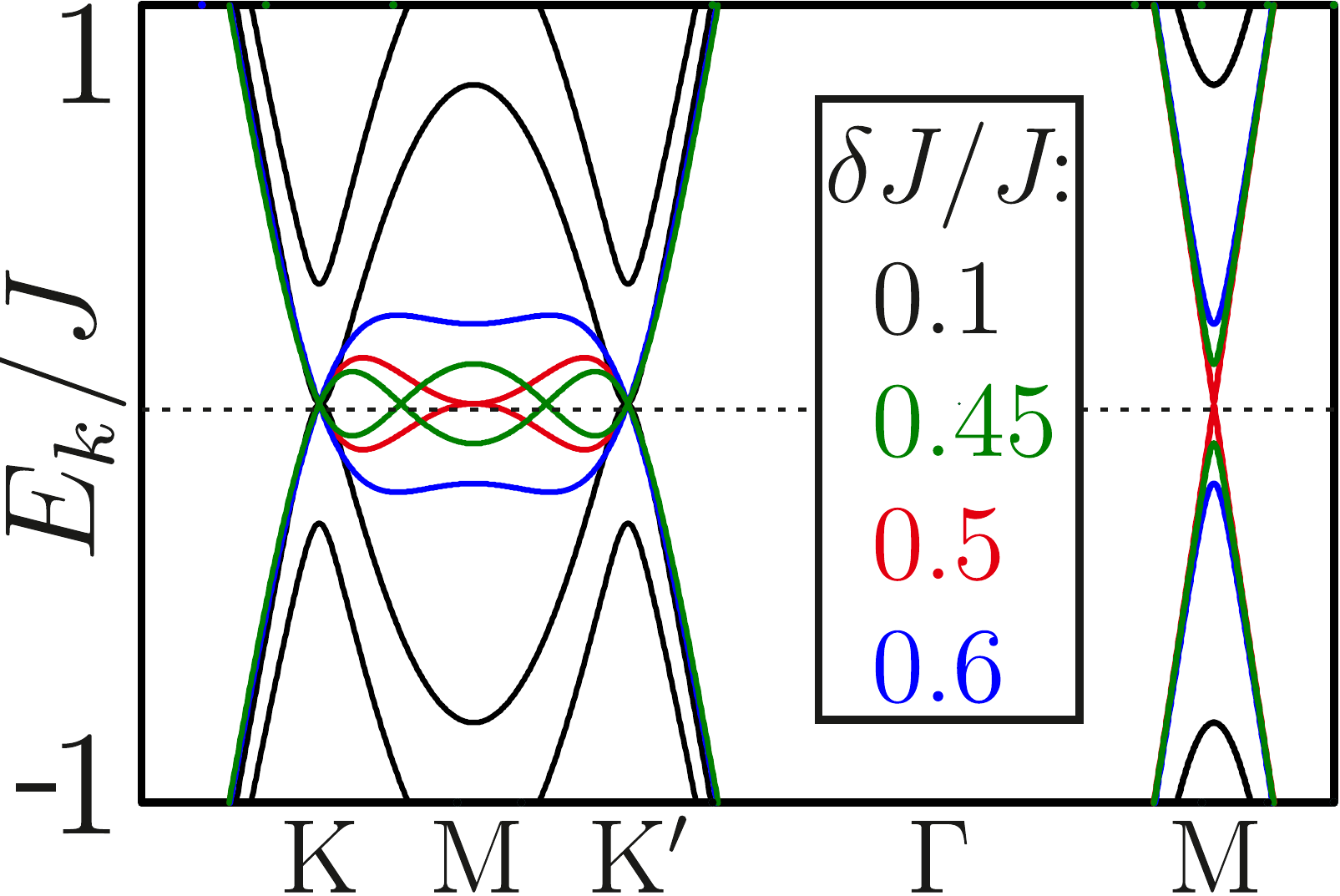} 
\caption{Topological transition sketch: additional Dirac touching points appear on the edges of the Brillouin zone due to TE-TM splitting in the absence of magnetic field. With increasing $\delta J$ they move away from the corners. At $\delta J=J/2$ additional Dirac points from neighbouring corners (K and K$^\prime$) meet in pairs at the centers of the edges (M). With further increase of $\delta J$, additional touching points of the dispersion disappear, leaving Dirac points only at K and K$^\prime$ points. }
\end{figure} 

The results of the main text were obtained for the most realistic case of rather small SOC ($\delta J / J = 0.1$).
In this case the dispersion topology in the absence of magnetic field is characterized by the trigonal warping effect, typical for monolayer graphene with Rashba spin-orbit interaction and bilayer graphene \cite{Rakyta2010,McCann2013}.
It consists in emergence of three additional Dirac cones in the vicinity of each Brilouin zone corner.
However, at a critical strength of the spin-orbit interaction $\delta J = J/2$, a transition occurs in the topology of the dispersion: additional Dirac cones with opposite Chern numbers meet in pairs at the centers of Brillouin zone edges (M points) and recombine.
This leads to a change of the band Chern number set from $C_n=\pm 2$ to $C_n=\pm 1$.

The transition is demonstrated in Figure S4 the in absence of the magnetic field.
Additional Dirac points are formed on the KK$^\prime$ edge of the Brioullin zone ($\delta J/J=0.1$, $\delta J/J=0.45$).
They meet at the M point and form linear energy dispersion in $\Gamma$M direction and a parabolic one in orthogonal KM direction ($\delta J/J=0.5$).
Then, the additional touching points of the dispersion disappear ($\delta J/J=0.6$).


\begin{thebibliography}{99}
\expandafter\ifx\csname url\endcsname\relax
  \def\url#1{\texttt{#1}}\fi
\expandafter\ifx\csname urlprefix\endcsname\relax\def\urlprefix{URL }\fi
\providecommand{\bibinfo}[2]{#2}
\providecommand{\eprint}[2][]{\url{#2}}

\bibitem{Klitzing1980}
\bibinfo{author}{Klitzing, K.~v.}, \bibinfo{author}{Dorda, G.} \&
  \bibinfo{author}{Pepper, M.}
\newblock \bibinfo{title}{New method for high-accuracy determination of the
  fine-structure constant based on quantized hall resistance}.
\newblock \emph{\bibinfo{journal}{Phys. Rev. Lett.}}
  \textbf{\bibinfo{volume}{45}}, \bibinfo{pages}{494--497}
  (\bibinfo{year}{1980}).

\bibitem{Kane2005}
\bibinfo{author}{Kane, C.~L.} \& \bibinfo{author}{Mele, E.~J.}
\newblock \bibinfo{title}{Quantum spin hall effect in graphene}.
\newblock \emph{\bibinfo{journal}{Phys. Rev. Lett.}}
  \textbf{\bibinfo{volume}{95}}, \bibinfo{pages}{226801}
  (\bibinfo{year}{2005}).

\bibitem{Haldane2008}
\bibinfo{author}{Haldane, F. D.~M.} \& \bibinfo{author}{Raghu, S.}
\newblock \bibinfo{title}{Possible realization of directional optical
  waveguides in photonic crystals with broken time-reversal symmetry}.
\newblock \emph{\bibinfo{journal}{Phys. Rev. Lett.}}
  \textbf{\bibinfo{volume}{100}}, \bibinfo{pages}{013904}
  (\bibinfo{year}{2008}).

\bibitem{Rechtsman2013}
\bibinfo{author}{Rechtsman, M.~C.} \emph{et~al.}
\newblock \bibinfo{title}{Photonic floquet topological insulators}.
\newblock \emph{\bibinfo{journal}{Nature}} \textbf{\bibinfo{volume}{496}},
  \bibinfo{pages}{196--200} (\bibinfo{year}{2013}).

\bibitem{Ozawa2014}
\bibinfo{author}{Ozawa, T.} \& \bibinfo{author}{Carusotto, I.}
\newblock \bibinfo{title}{Anomalous and quantum hall effects in lossy photonic
  lattices}.
\newblock \emph{\bibinfo{journal}{Phys. Rev. Lett.}}
  \textbf{\bibinfo{volume}{112}}, \bibinfo{pages}{133902}
  (\bibinfo{year}{2014}).

\bibitem{Jacqmin2014}
\bibinfo{author}{Jacqmin, T.} \emph{et~al.}
\newblock \bibinfo{title}{Direct observation of Dirac cones and a flatband in a
  honeycomb lattice for polaritons}.
\newblock \emph{\bibinfo{journal}{Phys. Rev. Lett.}}
  \textbf{\bibinfo{volume}{112}}, \bibinfo{pages}{116402}
  (\bibinfo{year}{2014}).

\bibitem{Leyder2007}
\bibinfo{author}{Leyder, C.} \emph{et~al.}
\newblock \bibinfo{title}{Observation of the optical spin Hall effect}.
\newblock \emph{\bibinfo{journal}{Nat Phys}} \textbf{\bibinfo{volume}{3}},
  \bibinfo{pages}{628--631} (\bibinfo{year}{2007}).

\bibitem{Nalitov2014}
\bibinfo{author}{Nalitov, A.~V.}, \bibinfo{author}{Solnyshkov, D.~D.},
  \bibinfo{author}{Tercas, H.} \& \bibinfo{author}{Malpuech, G.}
\newblock \emph{\bibinfo{journal}{arXiv:1404.6395}}  (\bibinfo{year}{2014}).

\bibitem{Fischer2014}
\bibinfo{author}{Fischer, J.} \emph{et~al.}
\newblock \bibinfo{title}{Anomalies of a nonequilibrium spinor polariton
  condensate in a magnetic field}.
\newblock \emph{\bibinfo{journal}{Phys. Rev. Lett.}}
  \textbf{\bibinfo{volume}{112}}, \bibinfo{pages}{093902}
  (\bibinfo{year}{2014}).

\bibitem{TKNN1982}
\bibinfo{author}{Thouless, D.~J.}, \bibinfo{author}{Kohmoto, M.},
  \bibinfo{author}{Nightingale, M.~P.} \& \bibinfo{author}{den Nijs, M.}
\newblock \bibinfo{title}{Quantized hall conductance in a two-dimensional
  periodic potential}.
\newblock \emph{\bibinfo{journal}{Phys. Rev. Lett.}}
  \textbf{\bibinfo{volume}{49}}, \bibinfo{pages}{405--408}
  (\bibinfo{year}{1982}).

\bibitem{Simon1983}
\bibinfo{author}{Simon, B.}
\newblock \bibinfo{title}{Holonomy, the quantum adiabatic theorem, and berry's
  phase}.
\newblock \emph{\bibinfo{journal}{Phys. Rev. Lett.}}
  \textbf{\bibinfo{volume}{51}}, \bibinfo{pages}{2167--2170}
  (\bibinfo{year}{1983}).

\bibitem{Kane2010}
\bibinfo{author}{Hasan, M.~Z.} \& \bibinfo{author}{Kane, C.~L.}
\newblock \bibinfo{title}{colloquium: Topological insulators}.
\newblock \emph{\bibinfo{journal}{Rev. Mod. Phys.}}
  \textbf{\bibinfo{volume}{82}}, \bibinfo{pages}{3045--3067}
  (\bibinfo{year}{2010}).

\bibitem{Novoselov2007}
\bibinfo{author}{Novoselov, K.~S.} \emph{et~al.}
\newblock \bibinfo{title}{Room-temperature quantum hall effect in graphene}.
\newblock \emph{\bibinfo{journal}{Science}} \textbf{\bibinfo{volume}{315}},
  \bibinfo{pages}{1379} (\bibinfo{year}{2007}).

\bibitem{Haldane1988}
\bibinfo{author}{Haldane, F. D.~M.}
\newblock \bibinfo{title}{Model for a quantum hall effect without landau
  levels: Condensed-matter realization of the "parity anomaly"}.
\newblock \emph{\bibinfo{journal}{Phys. Rev. Lett.}}
  \textbf{\bibinfo{volume}{61}}, \bibinfo{pages}{2015--2018}
  (\bibinfo{year}{1988}).

\bibitem{Kane2005b}
\bibinfo{author}{Kane, C.~L.} \& \bibinfo{author}{Mele, E.~J.}
\newblock \bibinfo{title}{Z2 topological order and the quantum spin hall
  effect}.
\newblock \emph{\bibinfo{journal}{Phys. Rev. Lett.}}
  \textbf{\bibinfo{volume}{95}}, \bibinfo{pages}{146802}
  (\bibinfo{year}{2005}).

\bibitem{Konig2007}
\bibinfo{author}{Konig, M.} \emph{et~al.}
\newblock \bibinfo{title}{Quantum spin hall insulator state in hgte quantum
  wells}.
\newblock \emph{\bibinfo{journal}{Science}} \textbf{\bibinfo{volume}{318}},
  \bibinfo{pages}{766--770} (\bibinfo{year}{2007}).

\bibitem{Hsieh2008}
\bibinfo{author}{Hsieh, D.} \emph{et~al.}
\newblock \bibinfo{title}{A topological dirac insulator in a quantum spin hall
  phase}.
\newblock \emph{\bibinfo{journal}{Nature}} \textbf{\bibinfo{volume}{452}},
  \bibinfo{pages}{970--974} (\bibinfo{year}{2008}).

\bibitem{Lindner2011}
\bibinfo{author}{Lindner, N.~H.}, \bibinfo{author}{Refael, G.} \&
  \bibinfo{author}{Galitski, V.}
\newblock \bibinfo{title}{Floquet topological insulator in semiconductor
  quantum wells}.
\newblock \emph{\bibinfo{journal}{Nat Phys}} \textbf{\bibinfo{volume}{7}},
  \bibinfo{pages}{490--495} (\bibinfo{year}{2011}).

\bibitem{Wang2008}
\bibinfo{author}{Wang, Z.}, \bibinfo{author}{Chong, Y.~D.},
  \bibinfo{author}{Joannopoulos, J.~D.} \& \bibinfo{author}{Solja\ifmmode
  \check{c}\else \v{c}\fi{}i\ifmmode~\acute{c}\else \'{c}\fi{}, M.}
\newblock \bibinfo{title}{Reflection-free one-way edge modes in a gyromagnetic
  photonic crystal}.
\newblock \emph{\bibinfo{journal}{Phys. Rev. Lett.}}
  \textbf{\bibinfo{volume}{100}}, \bibinfo{pages}{013905}
  (\bibinfo{year}{2008}).

\bibitem{Carusotto2012}
\bibinfo{author}{Umucalilar, R.~O.} \&
  \bibinfo{author}{Carusotto, I.}
\newblock \bibinfo{title}{Fractional quantum hall states of photons in an array
  of dissipative coupled cavities}.
\newblock \emph{\bibinfo{journal}{Phys. Rev. Lett.}}
  \textbf{\bibinfo{volume}{108}}, \bibinfo{pages}{206809}
  (\bibinfo{year}{2012}).

\bibitem{Hafezi2011}
\bibinfo{author}{Hafezi, M.}, \bibinfo{author}{Demler, E.~A.},
  \bibinfo{author}{Lukin, M.~D.} \& \bibinfo{author}{Taylor, J.~M.}
\newblock \bibinfo{title}{Robust optical delay lines with topological
  protection}.
\newblock \emph{\bibinfo{journal}{Nat Phys}} \textbf{\bibinfo{volume}{7}},
  \bibinfo{pages}{907--912} (\bibinfo{year}{2011}).

\bibitem{Khanikaev2013}
\bibinfo{author}{Khanikaev, A.~B.} \emph{et~al.}
\newblock \bibinfo{title}{Photonic topological insulators}.
\newblock \emph{\bibinfo{journal}{Nat Mater}} \textbf{\bibinfo{volume}{12}},
  \bibinfo{pages}{233--239} (\bibinfo{year}{2013}).

\bibitem{Chen2014}
\bibinfo{author}{Chen, W.-J.}, \bibinfo{author}{Jiang, S.-J.},
  \bibinfo{author}{Chen, X.-D.}, \bibinfo{author}{Dong, J.-W.} \&
  \bibinfo{author}{Chan, C.~T.}
\newblock \bibinfo{title}{Experimental realization of photonic topological
  insulator in a uniaxial metacrystal waveguide}.
\newblock \emph{\bibinfo{journal}{arXiv:1401.0367}}  (\bibinfo{year}{2014}).

\bibitem{Carusotto2013b}
\bibinfo{author}{Carusotto, I.} \& \bibinfo{author}{Ciuti, C.}
\newblock \bibinfo{title}{Quantum fluids of light}.
\newblock \emph{\bibinfo{journal}{Rev. Mod. Phys.}}
  \textbf{\bibinfo{volume}{85}}, \bibinfo{pages}{299--366}
  (\bibinfo{year}{2013}).

\bibitem{Sala2014}
\bibinfo{author}{Sala, V.~G.} \emph{et~al.}
\newblock \bibinfo{title}{Engineering spin-orbit coupling for photons and
  polaritons in microstructures}.
\newblock \emph{\bibinfo{journal}{arXiv:1406.4816}}  (\bibinfo{year}{2014}).

\bibitem{Kim2011}
\bibinfo{author}{Kim, N.~Y.} \emph{et~al.}
\newblock \bibinfo{title}{Dynamical d-wave condensation of exciton-polaritons
  in a two-dimensional square-lattice potential}.
\newblock \emph{\bibinfo{journal}{Nat Phys}} \textbf{\bibinfo{volume}{7}},
  \bibinfo{pages}{681--686} (\bibinfo{year}{2011}).

\bibitem{CerdaMendez2013}
\bibinfo{author}{Cerda-M\'endez, E.~A.} \emph{et~al.}
\newblock \bibinfo{title}{Exciton-polariton gap solitons in two-dimensional
  lattices}.
\newblock \emph{\bibinfo{journal}{Phys. Rev. Lett.}}
  \textbf{\bibinfo{volume}{111}}, \bibinfo{pages}{146401}
  (\bibinfo{year}{2013}).

\bibitem{suppl}
\bibinfo{note}{See Supplemental Material for extra results of numerical
  simulations.}

\bibitem{Shelykh2010}
\bibinfo{author}{Shelykh, I.~A.}, \bibinfo{author}{Kavokin, A.~V.},
  \bibinfo{author}{Rubo, Y.~G.}, \bibinfo{author}{Liew, T. C.~H.} \&
  \bibinfo{author}{Malpuech, G.}
\newblock \bibinfo{title}{Polariton polarization-sensitive phenomena in planar
  semiconductor microcavities}.
\newblock \emph{\bibinfo{journal}{Semiconductor Science and Technology}}
  \textbf{\bibinfo{volume}{25}}, \bibinfo{pages}{013001}
  (\bibinfo{year}{2010}).

\bibitem{Dasbach2005}
\bibinfo{author}{Dasbach, G.} \emph{et~al.}
\newblock \bibinfo{title}{Polarization inversion via parametric scattering in
  quasi-one-dimensional microcavities}.
\newblock \emph{\bibinfo{journal}{Phys. Rev. B}} \textbf{\bibinfo{volume}{71}},
  \bibinfo{pages}{161308} (\bibinfo{year}{2005}).

\bibitem{Galbiati2012}
\bibinfo{author}{Galbiati, M.} \emph{et~al.}
\newblock \bibinfo{title}{Polariton condensation in photonic molecules}.
\newblock \emph{\bibinfo{journal}{Phys. Rev. Lett.}}
  \textbf{\bibinfo{volume}{108}}, \bibinfo{pages}{126403}
  (\bibinfo{year}{2012}).


\bibitem{Rakyta2010} P. Rakyta, A. Kormanyos, and J. Cserti, Phys. Rev. B \textbf{82}, 113405 (2010).

\bibitem{McCann2013} Edward McCann and Mikito Koshino, Rep. Prog. Phys. 76 056503 (2013).
\end{thebibliography}
\end{document}